\documentclass[3p,authoryear]{elsarticle}
\usepackage{graphics}
\usepackage{graphicx}
\usepackage{amssymb}
\usepackage[colorlinks]{hyperref}

\journal{Computational Statistics \& Data Analysis}

\usepackage{hyperref}
\usepackage{amsmath}

\usepackage{color}  % nur um Kommentare hervorzuheben

\newcommand{\med}{\mathop{\mbox{med}}}

\begin{document}

\begin{frontmatter}

%% Title, authors and addresses

%% use the tnoteref command within \title for footnotes;
%% use the tnotetext command for theassociated footnote;
%% use the fnref command within \author or \address for footnotes;
%% use the fntext command for theassociated footnote;
%% use the corref command within \author for corresponding author footnotes;
%% use the cortext command for theassociated footnote;
%% use the ead command for the email address,
%% and the form \ead[url] for the home page:
%% \title{Title\tnoteref{label1}}
%% \tnotetext[label1]{}
%% \author{Name\corref{cor1}\fnref{label2}}
%% \ead{email address}
%% \ead[url]{home page}
%% \fntext[label2]{}
%% \cortext[cor1]{}
%% \address{Address\fnref{label3}}
%% \fntext[label3]{}

\title{Cellwise Robust M Regression\footnote{Article has appeared as: Computational Statistics and Data Analysis, 147 (2020), 106944, DOI: \url{https://doi.org/10.1016/j.csda.2020.106944}}\footnote{\textcopyright\  2020. This manuscript version is made available under the CC-BY-NC-ND 4.0 license \url{http://creativecommons.org/licenses/by-nc-nd/4.0/}}}
%% use optional labels to link authors explicitly to addresses:
%% \author[label1,label2]{}
%% \address[label1]{}
%% \address[label2]{}
    
\author[tuw]{P. Filzmoser\corref{cor1}}
\ead{P.Filzmoser@tuwien.ac.at}
\author[kul]{S. H\"{o}ppner}
\ead{sebastiaan.hoppner@kuleuven.be}
\author[ast]{I. Ortner}
\ead{irene.ortner@applied-statistics.at}
\author[asp]{S. Serneels}
\ead{Sven.Serneels@aspentech.com}
\author[kul]{T. Verdonck}
\ead{tim.verdonck@kuleuven.be}
\cortext[cor1]{Corresponding author. Institute of Statistics and Mathematical Methods in Economics,
    TU Wien, Wiedner Hauptstra\ss e 8-10, 1040 Vienna, Austria. 
    Tel.: +43 1 58801 10560, Fax: +43 1 58801 105699}

\address[tuw]{Institute of Statistics and Mathematical Methods in Economics,
    TU Wien, Wiedner Hauptstra\ss e 8-10, 1040 Vienna, Austria}
\address[kul]{Department of Mathematics, K.U. Leuven, Leuven, Belgium}
\address[ast]{Applied Statistics GmbH, Vienna, Austria}
\address[asp]{Aspen Technology, Bedford, Massachusetts, MA01730, USA}

\begin{abstract}
The cellwise robust M regression estimator is introduced as the first estimator of its kind that intrinsically yields both a map of cellwise outliers consistent with the linear model, and a vector of regression coefficients that is robust against vertical outliers and leverage points. As a by-product, the method yields a weighted and imputed data set that contains estimates of what the values in cellwise outliers would need to amount to if they had fit the model. The method is illustrated to be equally robust as its casewise counterpart, MM regression. The cellwise regression method discards less information than any casewise robust estimator. Therefore, predictive power can be expected to be at least as good as casewise alternatives. These results are corroborated in a simulation study. Moreover, while the simulations show that predictive performance is at least on par with casewise methods if not better, an application to a data set consisting of compositions of Swiss nutrients, shows that in individual cases, CRM can achieve a much higher predictive accuracy compared to MM regression. 
\end{abstract}

\begin{keyword}
Cellwise Robust Statistics \sep Cellwise Robust M regression \sep Cellwise Outliers \sep Detecting Deviating Cells \sep Linear Regression

%% PACS codes here, in the form: \PACS code \sep code

%% MSC codes here, in the form: \MSC code \sep code
%% or \MSC[2008] code \sep code (2000 is the default)
\end{keyword}

\end{frontmatter}

%% \linenumbers

%% main text
\section{Introduction}

Linear regression is one of the most frequently studied problems in the statistical sciences. It is well known that the least squares estimator fulfill several optimality criteria under normal distribution assumptions, a result that goes all the way back to Gau\ss \  \citep{Gauss}. Likewise, it is well known that the least squares estimator is not optimal when data deviate from these assumptions. A lot of attention has been attributed to developing methods that still yield sensible regression parameters in the presence of  casewise deviations. Such casewise deviations may originate from a fraction $\epsilon$ of the data having been generated from a different distribution (outliers), or the data satisfying the linear model with a non-normal error term, such as a Cauchy or Student's $t$. In these cases, robust linear regression methods generally outperform their least squares counterpart. Many different approaches to casewise robust regression have been proposed, a good overview of which can be found in reference works \citet{Huber, MaronnaMY06,maronna2019robust,RousseeuwLeroy}. 

In the bulk of the literature on robust statistics, robustness is considered to be robustness against entire cases that do not satisfy model assumptions. For a univariate predictor $\mathbf{x}=(x_1,\ldots ,x_n)^T$, this approach is plausible because it corresponds to individual elements $x_i$ either fitting the assumptions or not. Conversely, assuming that outliers are complete observations of a multivariate predictor,
thus multivariate observations where each
\textcolor{black}{entry in the observation vector}
is considered as an 
outlier, may not correspond to reality. In real life, the 
predictor matrix often consists of single predictors
that are measurements of different physical entities, which need not generate outliers simultaneously. Imagine, for example, each column being a sensor in a manufacturing plant. Whereas it is viable to assume multivariate interplay between these sensors to be present under normal operating conditions, each of these sensors may break down independently and therefore, generate outliers individually. Another example would be gene expression in microarray data, and there are many more. Discarding whole cases in these (and other) practical situations can cause a significant loss of information in the estimation procedure, which just like harsh downweighting of entire outliers, can be surmised to increase estimation variance.

In the light of the above, to make maximal use of the non-contaminated portion of the data, in practice it is often preferable to detect outliers on a cellwise basis instead of casewise. 
\textcolor{black}{This means that single entries (cells) in the data matrix are 
considered as potential outliers, and not necessarily a whole row (observation).}
Up to today, this usually implies that outlier detection is done as a separate step {\em before} the remainder of the analysis. However, any outlier is only outlying with respect to a model and therefore, such a preliminary outlier detection c.q. correction step may distort the data in a way that is inconsistent with the model. There is a large gap yet to be covered in method development on cellwise robust techniques: methods that allow to detect and correct for deviating cells in a single model consistent way. Cellwise robust regression is still a nascent field of research. In this paper, a new cellwise robust M regression estimator (CRM) is proposed. In one run, it allows to estimate regression coefficients that are robust against cellwise and casewise outliers, while also providing a map of the deviating cells. The option to construct the estimator as a cellwise robust M regression as opposed to alternative paths, such as 
MCD regression \citep{Rousseeuw84}, comes from the observation that robust M regression estimators have proven to yield a very good trade-off between efficiency and robustness in simulations and applications in fields as diverse as quantitative structure--property relationships (QSPR) \citep{SS-PP}, gravimetry \citep{gravimeters},
\textcolor{black}{finance} \citep{Guerard}, 
chemometrics \citep{SPRM}, analytical chemistry with applications to e.g. analysis of arch\ae ological glass \citep{PRM} and meteorite samples \citep{SPRM-DA}, as well as estimation of shaping coefficients for futures trading in the electricity markets \citep{MCRM}. Note though, that S-regression has also proven a valid path in this context \citep{Oellerer_Shooting}. 

Motivated by this assumption, in this manuscript the {\em cellwise robust M} (CRM) regression estimator is introduced. It consists of an iteratively reweighted least squares procedure, starting with weights derived from highly robust estimates, that both compensate for casewise vertical outliers and leverage points. Within each iteration, the SPADIMO \citep{SPADIMO} procedure is applied, detecting the cells that contribute most to outlyingness. The re-weighting scheme is then adapted to only downweight outlying cells. The resulting method thereby can deliver a highly robust estimate of regression coefficients (and intercept), and in a model consistent way, yield cellwise outlier detection. Because not as much information in the data is discarded, the method should be more efficient than a casewise robust estimator.   

The article is organized as follows. In Section \ref{sec:Algo}, the CRM algorithm is described in detail. Section \ref{sec:Simulations} presents a simulation study comparing CRM to different approaches in terms of efficiency, as well as in terms of its capability to detect and downweight the correct set of outlying cells. In Section \ref{sec:Example}, the method is applied to a compelling example. Finally, Section \ref{sec:Outro} concludes. 

\section{The CRM algorithm} \label{sec:Algo}

\subsection{Background}
The target of this section is to propose an estimator for the linear model that is robust against cellwise outliers, and as a by-product, yields a map of the detected outlying cells. 

Let $\mathbf{X} \in \mathbb{R}^{n \times p}$ be a predictor matrix consisting of $n$ cases of $p$ predictor variables (or, if an intercept is considered, $p-1$ predictors, and the first column with ones for the intercept) and let $\boldsymbol{\beta} \in \mathbb{R}^p$ be a fixed, true vector of regression coefficients. Then, in the linear model, $n$ cases of a univariate dependent variable  $\mathbf{y}\in \mathbb{R}^n$ 
relate to the predictors as 
\begin{equation}\label{eq:linmod}
\mathbf{y} = \mathbf{X} \boldsymbol{\beta} + \boldsymbol{\varepsilon},
\end{equation}
where the entries of $\boldsymbol{\varepsilon}$ are independent and identically distributed (i.i.d.) and where $E(\boldsymbol{\varepsilon}) = \mathbf{0}$ and $Cov\left(\boldsymbol{\varepsilon}\right) = \sigma^2 \mathbf{I}_n$. As signaled before, while the least squares estimator may be optimal under normality assumptions, robust regression methods should be the estimators of choice when outliers are expected to be in the data. Several classes of robust regression estimators exist \citep[see, e.g.,][]{RousseeuwLeroy,MaronnaMY06}). Performance analysis of robust estimators has several facets, but most prominently it comes down to analyzing how well the estimators perform in trading off robustness for statistical efficiency. Estimators that can resist a high fraction of outliers in the data, tend to have a higher variance than the corresponding maximum likelihood estimator, but that loss in efficiency need not be dramatic. Along other classes of methods, MM estimators 
\citep{Yohai87} are known to perform well in terms of the robustness--efficiency trade-off and have, for that reason, been incorporated into mainstay implementations of robust regression, such as the function {\sf lmrob()} in the {\sf R} package {\sf robustbase} \citep{robustbase}. 

To understand how MM estimators work, let us revert to least squares. By definition, least squares minimizes a loss function of squared residuals. 
Consider a given estimator $\hat{\boldsymbol{\beta}}$ of $\boldsymbol{\beta}$. Then the $i$-th regression residual is defined as
$r_i(\hat{\boldsymbol{\beta}})=y_i - \mathbf{x}_i^T\hat{\boldsymbol{\beta}}$, where $\mathbf{y}=(y_1,\ldots ,y_n)^T$ and $\mathbf{x}_i$ represents the $i$-th row of the data 
matrix $\mathbf{X}$, for $i=1,\ldots ,n$.
The least squares estimator of the regression coefficients is
given by the minimization problem
\begin{equation}\label{critLS}
    \hat{\boldsymbol{\beta}}_{\mathrm{LS}} = \mathop{\mbox{argmin}}_{\boldsymbol{\beta}}\sum_i{r_i(\boldsymbol{\beta})^2}.
\end{equation}
A more general definition is 
\begin{equation}\label{critM}
    \hat{\boldsymbol{\beta}} = \mathop{\mbox{argmin}}_{\boldsymbol{\beta}}\sum_i
    \rho\left(\frac{r_i(\boldsymbol{\beta})}{\hat{\sigma}}\right) ,
\end{equation}
where $\hat{\sigma}$ is a robust scale estimator of the residuals, and 
$\rho(r)$ is a function that is approximately quadratic for small (absolute) $r$,
but increases more slowly than $r^2$ for larger values of $r$. Moreover,
the inclusion of $\hat{\sigma}$ allows to get the same result if the response is
rescaled. This definition (\ref{critM}) is referred to as the class of M-estimators \citep{Huber}.
Not all choices of $\rho$ have practical relevance, but when diligently chosen, M-estimators can have a bounded influence regarding deviating, or even erratic values in
the response, and therefore they can be robust against vertical outliers. 
The robustness properties of the resulting estimator derive back to $\rho$, or more precisely, to its derivative $\psi=\rho '$. More particularly, common practicable choices leading to robust estimators for $\psi$, such as the Huber, Hampel redescending or Hampel-Rousseeuw hyperbolic tangent functions \citep{Hampel}, depend on a set of parameters. The resulting robustness properties then become a function of the corresponding parameters. In this paper, the Hampel redescending function is chosen, with a similar motivation as to why the MM estimation path is pursued. The Hampel function, in its reweighting representation, is given by: 
\begin{equation}
\label{eq:hampel}
w_H(r) = \left\{\begin{matrix}%{ccc} 
1 & & |r| \leq Q_1 \\
\frac{Q_1}{|r|} & & Q_1 < |r| \leq Q_2 \\
\frac{Q_3 - r}{Q_3 -Q_2} \frac{Q_1}{|r|} & \mbox{if} & Q_2 < |r| \leq Q_3 \\
0 & & Q_3 < |r|\end{matrix} 
 \right. .
\end{equation}
It depends on a set of three parameters $Q_1$, $Q_2$ and $Q_3$. When applied to regression residuals, which for standardized data can be assumed to be standard normally distributed, sensible values for the parameters are the $0.95$, $0.975$ and $0.999$ quantiles of the standard normal distribution. 

It can be shown that in practice, calculating M-estimators directly through optimizing \eqref{critM} is equivalent to running an iteratively re-weighted least squares (IRLS) procedure \citep{Green}. However, the resulting estimator will in general only be 
robust against outliers in the response, i.e.~vertical outliers. In order to
achieve robustness also against outliers in the explanatory variables (leverage points),
it is important to select a robust starting estimator $\hat{\boldsymbol{\beta}}_0$.
This can be done by taking a highly robust but inefficient S-estimator \citep{RY1984},
combined with a robust M-scale estimator $\hat{\sigma}$ \citep{Huber}. The resulting robust
{\em MM} estimator inherits the $50\%$ breakdown point of the S-estimator, and
has tunable efficiency \citep[see][for more details]{MaronnaMY06}.

In this paper, robust MM estimators are now being generalized to cellwise robustness. In order to achieve this, the IRLS procedure will need a way to be able to detect which cells are outlying. Exactly for this purpose, the method of Sparse Directions of Maximal Outlyingness (SPADIMO) \citep{SPADIMO} has recently been developed. SPADIMO is a method that identifies which variables contribute most to a case being detected as an outlier. By incorporating SPADIMO into the IRLS reweighting scheme, and only downweighting cells flagged by SPADIMO, the method will be cellwise robust.  

\subsection{The Algorithm}

The overarching algorithm described in this section can be seen as a way to convey cellwise robustness properties to any given robust regression method. However, robust regression methods being significantly different by construction, many details in the algorithm need to be adapted to the specific regression method. As outlined before, we have opted to develop the algorithm specifically as a cellwise extension to robust MM regression, inspired by the good robustness--efficiency tradeoff which have been shown in theory, simulation and practical 
applications \citep{MaronnaMY06}.

MM regression estimators consist of two steps: at first, a highly robust initial estimate is calculated, which conveys its high breakdown point to the entire procedure. Then, the highly robust initial estimate is used as a plug-in estimator for an M-estimator, which in practice means that the initial estimate is used as a starting point for an algorithm to achieve higher efficiency by iterative reweighting. How this concept can be used to obtain an efficient and highly robust cellwise regression method is presented below in detail, and an overview of the essential steps is given in Algorithm 1. 

\paragraph{Initial outlier detection}

Prior to starting the algorithm, the data should be centered and scaled. Since the data may still contain both cellwise and casewise outliers at that point, the preprocessing should be done robustly, with estimators that have a 50\% asymptotic breakdown point. Good choices for centering and scaling the data robustly would be the $L_1$ median and the $Q_n$ scale estimator \citep{RousseeuwQn}, respectively, but viable alternatives to these choices exist. Several algorithms to compute the $L_1$ median are available, and a good comparison is given in \citet{L1median_algos}. 

In the spirit of MM estimation, an initial, highly robust regression estimator is used to identify suspected casewise outliers. Here we use the MM estimator as starting point, but a good alternative would be the LTS estimator, which is also known to be an efficient and robust regression estimator \citep{fastLTS}.
Based on the robust MM regression estimator, observations are flagged as casewise outliers if their absolute standardized residuals exceed the 95\% quantile of the standard normal distribution. For each of the casewise outliers, it now has to be determined if they truly are casewise outliers, or if there is a subset of cells in them which make them outlying. In order to investigate which variables contribute most to the outlyingness of the casewise outliers, the SPADIMO \citep{SPADIMO} algorithm is applied. For those cases that contain cellwise outliers, outlying variables are imputed as if they were missing cells (see Algorithm 2). To impute the values in the outlying cells, the two nearest neighbors are detected based on the {\em clean} cells in the case. This means that, when a set $\mathcal{C}$ of $q < p$ variables have been detected as cellwise outliers in the case, the two nearest neighbors are determined in the $\mathbb{R}^{p-q}$ variate space spanned by the variables in $\{1,\ldots,p\} \backslash \mathcal{C}$. The nearest neighbor search is only carried out in the subset of cases that are not outlying, i.e.~have case weights equal to one. The cellwise outliers in the case under consideration are now being imputed with the corresponding column means of the two nearest neighbors. This imputation procedure generates modified cases with smaller residuals, which increases their case weights and by consequence, the valuable information in the non-outlying variables contributes to the model. This imputation step is not only a part of the algorithm initiation, but will also take place in each IRLS step. As such, the initial outlier detection step yields a first value of estimates for the regression coefficients $\hat{\boldsymbol{\beta}}$, as well as a first set of weighted and imputed data $\mathbf{X}_\omega$ and $\mathbf{y}_\omega$ for the explanatory
variables and the response, respectively.

\paragraph{Iterative modeling and outlier detection.}

Once an initial highly cellwise robust estimate of the regression coefficients has been obtained, a more efficient estimate can be found by using this estimate as a starting value into an iteratively reweighted least squares (IRLS) routine. Starting from the initial $\hat{\boldsymbol{\beta}}$, $\mathbf{X}_\omega$ and $\mathbf{y}_\omega$, the first IRLS update will consist of least squares regression estimates 
based on the weighted data $\mathbf{X}_\omega$ and $\mathbf{y}_\omega$. In each step, the residuals are calculated, and casewise outliers are detected based on the magnitude of the residuals. For those cases flagged as outliers, variables contributing to outlyingness are found by SPADIMO, and for the cases that are not entirely outlying, the outlying cells are imputed as before. Now the residuals can be recalculated based on the newly imputed data and a new set of weights is computed. The procedure continues until 
the estimated regression coefficients stabilize. 

The complete algorithm is as follows:
\begin{itemize}
\item Apply robust regression (e.g.~MM regression) on the original 
observations $\mathbf{x}_i$
and $y_i$, for $i=1,\ldots ,n$, to obtain the initial estimator $\hat{\boldsymbol{\beta}}$.
\item Run Algorithm 1, starting with the original observations
and the initial regression estimator.
\item Run Algorithm 1, starting with the resulting weighted and imputed
data $\mathbf{X}_{\omega}$ and $\mathbf{y}_{\omega}$ from the previous 
point, and with
the least squares estimator $\hat{\boldsymbol{\beta}}$ from a regression
using these data.
\item Run Algorithm 1 with the weighted data as a result from the previous 
point, and the corresponding least squares estimator, and repeat
until the mean absolute difference of the subsequent regression estimates
is smaller than a tolerance bound (e.g.~0.01).
\end{itemize}
The {\sf R} code implementations of CRM and SPADIMO are available in the {\sf R} package {\sf crmReg} at 
\begin{center}
github.com/SebastiaanHoppner/CRM.
\end{center}

\begin{minipage}{\textwidth}
\noindent\makebox[\linewidth]{\rule{\textwidth}{0.4pt}}
\textbf{ Algorithm 1:} CRM Iteratively reweighted least squares algorithm.

\vspace{-2mm}
\noindent\makebox[\linewidth]{\rule{\textwidth}{0.4pt}}

%\vspace{-2mm}
\begin{enumerate}
\item Calculate residuals based on the estimator $\hat{\boldsymbol{\beta}}$:
\begin{equation*}
r_i= y_i - \mathbf{x}_i^T \hat{\boldsymbol{\beta}} \quad \text{ for } i \in \{1,...,n\}
\end{equation*}
\item Detect outliers as cases that satisfy
\begin{equation*}
     \frac{|r_i|}{c \med_j |r_j|} > z_{0.95} ,
\end{equation*}
where $c=1.4826$ for consistency of the MAD, and $z_{0.95}$ is the 0.95 quantile 
of the standard normal distribution.
\item For each outlying case:
\begin{itemize}
\item Apply SPADIMO and obtain outlying variables.
\item If not all variables contribute to outlyingness: Impute values in outlying variables as in Algorithm 2. 
\item Denote the newly imputed data matrix by $\tilde{\mathbf{X}}$.
\end{itemize}
\item Update residuals 
\begin{equation*}
\tilde{r}_i= y_i - \tilde{\mathbf{x}}_i^T \hat{\boldsymbol{\beta}} \quad \text{ for } 
i =1,...,n.
\end{equation*}
\item Calculate case weights by the Hampel weight function (\ref{eq:hampel}),
\begin{equation*}
    \omega_i = w_H\left( \frac{|\tilde{r}_i|}{c \med_j |\tilde{r}_j|} \right)
\end{equation*}
with $c$ as in Step 2.
\item 
\textcolor{black}{
Let $\boldsymbol{\Omega}=\mbox{Diag}(\sqrt{\omega_1},\ldots ,\sqrt{\omega_n})$ be a diagonal matrix with the case weights as diagonal elements. 
}
Update the (imputed) data as
\begin{equation*}
\mathbf{X}_{\omega} =\boldsymbol{\Omega} \tilde{\mathbf{X}}
\qquad \mbox{ and } \qquad
\mathbf{y}_{\omega} =\boldsymbol{\Omega} \mathbf{y} .
\end{equation*}
\end{enumerate}
\noindent\makebox[\linewidth]{\rule{\textwidth}{0.4pt}}
\text{}
\end{minipage}

\begin{minipage}{\textwidth}
\noindent\makebox[\linewidth]{\rule{\textwidth}{0.4pt}}
\textbf{ Algorithm 2:} CRM imputation algorithm.

\vspace{-2mm}
\noindent\makebox[\linewidth]{\rule{\textwidth}{0.4pt}}

\begin{enumerate}
    \item Let $i$ be the index of an outlying case ${\mathbf{x}}_i$.
    \item Let $\mathcal{C}$ be the set of $q < p$ variables detected as cellwise outliers in ${\mathbf{x}}_i$.
    \item Detect the two nearest neighbors ${\mathbf{x}}_{k_1}$ and ${\mathbf{x}}_{k_2}$ of the outlier ${\mathbf{x}}_i$ in the subspace $\{1,\cdots,p\} \backslash \mathcal{C}$ and only among observations ${\mathbf{x}}_j$ with $\omega_j = 1$.
    \item Impute outlying cells $\tilde{x}_{iq} = ({x}_{k_1q} + {x}_{k_2q})/2$ with $q\in\mathcal{C}$.
\end{enumerate}
\noindent\makebox[\linewidth]{\rule{\textwidth}{0.4pt}}
\text{}
\end{minipage}

\section{Simulation study} \label{sec:Simulations}

Cellwise robust estimation is a fairly recent development in the statistical sciences. Up to our knowledge, there is no report of a cellwise robust M-type regression estimator. With an emphasis on cellwise outlier detection, the {\em Detecting Deviating Data Cells} (DDC) method has been proposed \citep{Rousseeuw_DDC}. At this point, it is noted that DDC has been designed with the purpose to yield reliable cellwise outlier detection, {\em even when $\mathit{> 50\%}$ of the cases contain outlying cells}. The CRM method proposed here will not be robust against contamination of more than half of the data. While it does not offer the latter advantage, it does yield model consistent cell weights in combination with an increased statistical efficiency when compared to casewise robust regression methods. 

In this simulation study, the performance of CRM applied to the robust coefficient estimator of an MM regression is compared to conventional MM regression, MM regression combined with DDC, ordinary least squares (OLS) regression and OLS regression combined with DDC. The simulation study establishes that CRM, as a method that intertwines cellwise robustness properties with estimating regression coefficients for the linear model in a model consistent way, significantly outperforms application of a model agnostic detection method for cellwise outliers (DDC), followed by either a classical or robust regression estimator.

\subsection{Simulation setting}

The data for the simulation study are generated from a $p$-dimensional multivariate normal distribution with center $\boldsymbol{\mu}=(0,\dots,0)^T$ and covariance matrix $\boldsymbol{\Sigma}$. The covariance matrix is a matrix of zeros, with ones in the diagonal and 0.5 in the first off-diagonal, so $\boldsymbol{\Sigma}_{i, i}=1$ for $i=1, \ldots, p$, $\boldsymbol{\Sigma}_{j, j+1}=\boldsymbol{\Sigma}_{j+1, j}=0.5$ for $j=1, \ldots, p-1$ and $\boldsymbol{\Sigma}$ is zero elsewhere. The number of variables is set to $p=50$ and $n=400$ cases are generated, resulting in the data matrix $\mathbf{X}\in\mathbb{R}^{n\times p}$.

Let $\boldsymbol{\beta}$ be a vector of length $p$ of random values from a standard normal distribution, normalized to length 10 and the intercept $\beta_0 = 10$. The error term $\boldsymbol{\epsilon}$ is a vector of length $n$ of random values from a normal distribution with mean 0 and standard deviation 0.5.

\begin{figure}[t]
\centering
\includegraphics[width=0.6\textwidth]{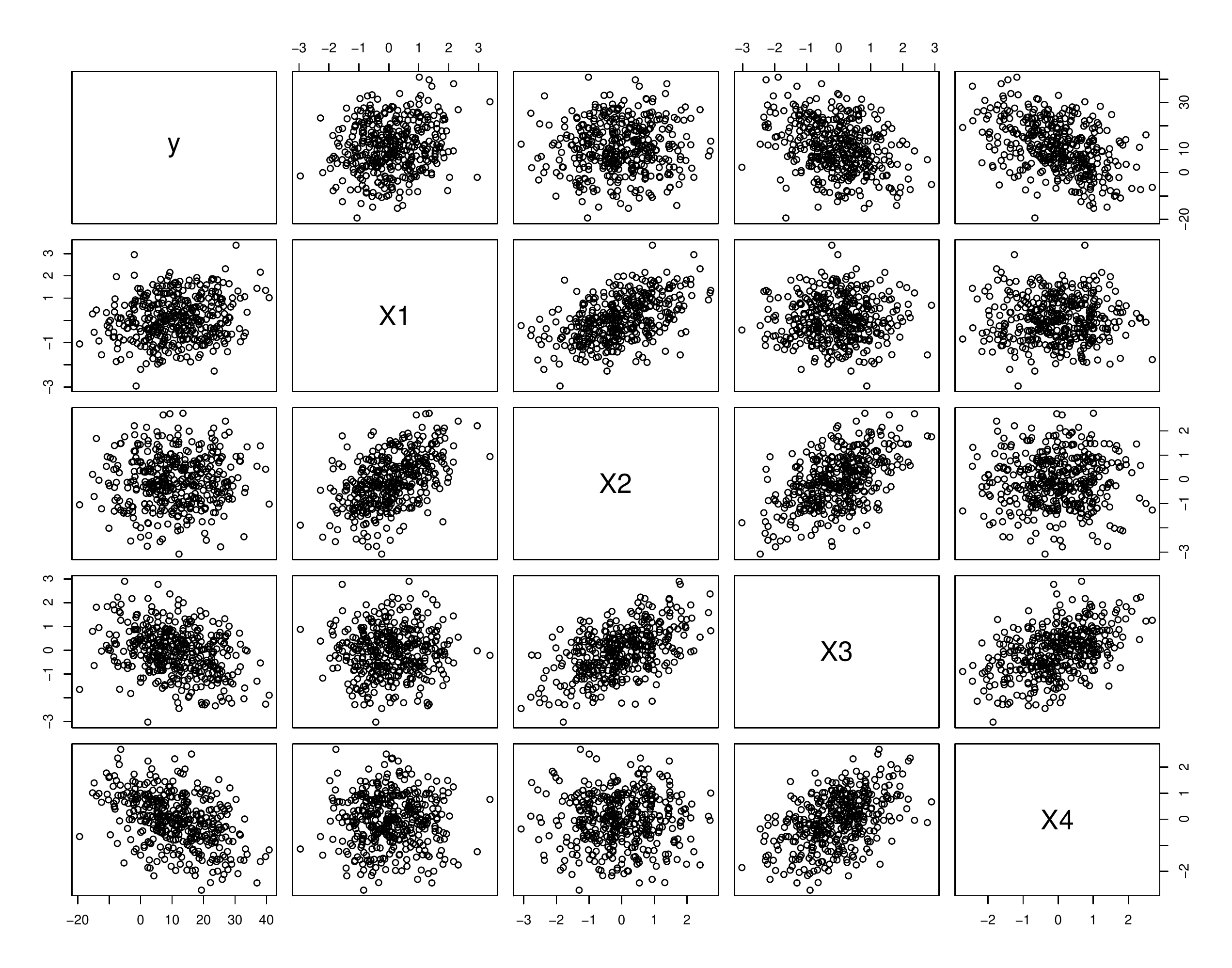}
~
\vspace*{-5mm}
\caption{Pairwise scatterplot of the response variable $\mathbf{y}$ and the first four predictor variables in $\mathbf{X}$.}
\label{fig:pairs_clean_data}
\end{figure}

Then, the response is generated for clean data as follows:
\begin{equation}
    \mathbf{y} = \mathbf{1}_n\beta_0 + \mathbf{X}  \boldsymbol{\beta} + \boldsymbol{\epsilon},
\end{equation}
so the clean data consists of $(\mathbf{y}, \mathbf{X})$ and the regression coefficients are $(\beta_0, \boldsymbol{\beta})$. A pairwise scatterplot of the response variable 
$\mathbf{y}$ and the first four predictor variables in $\mathbf{X}$ is shown in Figure \ref{fig:pairs_clean_data}.

\subsection{Adding contamination}
\label{sec:contam}

Contamination is added to the data matrix $\mathbf{X}$, and the contaminated matrix is denoted as $\mathbf{X}^c$. For the contamination we randomly select a fraction of $r=5\%$ of the observations in $\mathbf{X}$, so $r \cdot n=20$ rows of $\mathbf{X}$ are randomly selected. These $20$ observations will be contaminated and are called casewise outliers. 
\textcolor{black}{
Here, the fraction of contamination is fixed; note that in Section~\ref{sec:breakdown}, the effect of the fraction of contamination will be investigated.
}
Let $I^c\subset\{1, \ldots, n\}$ denote the random subset of 20 selected case indices. To generate cellwise outliers, for each selected case $i\in I^c$, $\check{r}=10\%$ of the predictor variables are randomly picked. So for each casewise outlier, $\check{r} \cdot p=5$ randomly selected cells will be contaminated. For each $i\in I^c$, let $J_i^c\subset\{1, \ldots, p\}$ denote the subset of 5 selected variable indices. The total number of contaminated cells in $\mathbf{X}^c$ will be $r\cdot \check{r}\cdot n\cdot p=100$.

\begin{figure}[t]
	\centering
	\includegraphics[width=0.6\textwidth]{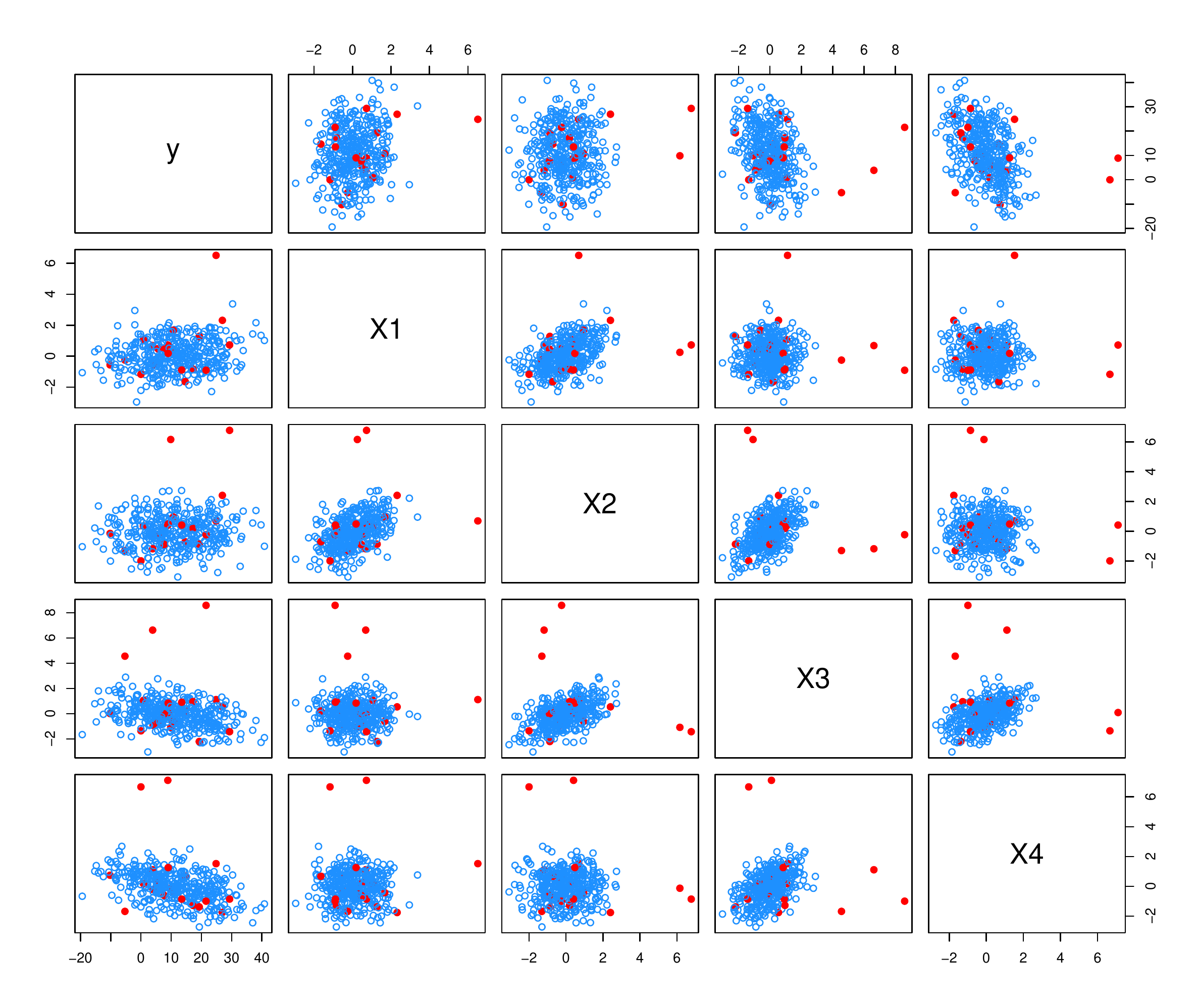}
	~
\vspace*{-5mm}
	\caption{Pairwise scatterplot of the response variable $\mathbf{y}$ and the first four predictor variables in $\mathbf{X}^c$. The casewise outliers are in red and the uncontaminated cases are in blue.}
	\label{fig:pairs_contaminated_data}
\end{figure}

Cellwise contamination in variable $j$ is achieved by adding to its mean value $\overline{x}_j$, $k=6$ times the standard deviation $s_j$ of variable $j$ plus a random value $e$ of the standard normal distribution. The contaminated matrix is $\mathbf{X}^c$ with
\[x^c_{ij}=\overline{x}_j + ks_j + e = \overline{x}_j + k\sqrt{\frac{1}{n-1}\sum_{l=1}^{n}(x_{lj} - \overline{x}_j)^2} + e\]
for all $i\in I^c$ and $j\in J^c_i$. The contaminated data consists of $(\mathbf{y}, \mathbf{X}^c)$ where a casewise outlier is considered an observation which has contaminated cells. Figure \ref{fig:pairs_contaminated_data} shows a pairwise scatterplot of the response variable $\mathbf{y}$ and the first four predictor variables in $\mathbf{X}^c$. The casewise outliers are in red and the uncontaminated cases are in blue.

\textcolor{black}{
Section~\ref{sec:varyk} will also contain simulation results where $k$ is varied within a certain range.
}

\subsection{Regression methods}

The following linear regression methods have been fit to the contaminated data $(\mathbf{y}, \mathbf{X}^c)$: CRM (with MM regression as a starting estimate), simple MM regression and OLS regression. The following parameter settings for the CRM regression estimation were used: the maximal number of iterations was set to 100, the relative tolerance for converging the regression coefficients was set to 0.01, the outlyingness factor for SPADIMO was set to 1.5; SPADIMO sparsity was allowed to vary as $\eta \in \{0.1,0.2,\cdots,0.9\}$ \citep{SPADIMO}. The authors  also suggest these settings as the default values in the \texttt{R} implementation. 

For the methods that are sequential combinations of DDC with regression methods, the workflow goes as follows. At first, the Detect Deviating Cells (DDC) method is applied to the contaminated data matrix $\mathbf{X}^c$ , which returns a DDC-imputed matrix, further denoted $\mathbf{X}^{DDC}$. Then, MM and OLS regression are fit to $(\mathbf{y}, \mathbf{X}^{DDC})$.

\subsection{Evaluation}

Four different criteria are assessed to evaluate the relative performance of the three approaches. 

At first, the most obvious performance criterion for any regression method is predictive performance on independent test data. To evaluate the prediction performance, the mean squared error of prediction (MSEP) is calculated over the set of uncontaminated cases:
\begin{equation}\label{eq:msep}
\text{MSEP}= \frac{1}{n_{clean}}\sum_{i \in I}(\hat{y}_i-y_i)^2
\end{equation}
where $I$ contains the indices of clean, uncontaminated cases ($I$ is the complementary set of $I^c$) and $n_{clean}$ is the number of uncontaminated cases.

Secondly, it is interesting to know how much the individual regression coefficients, estimated by the three methods, deviate from the truth. To assess bias for the individual regression coefficients, the mean absolute error (MAE)
\begin{equation}\label{eq:mae}
\text{MAE}=\frac{1}{p}\sum_{j=1}^{p}|\hat{\beta}_j-\beta_j|
\end{equation}
is reported. 

CRM and DDC each generate an imputed matrix of $\mathbf{X}^c$, denoted as $\mathbf{X}^{imp}$. In the case of DDC, $\mathbf{X}^{imp}=\mathbf{X}^{DDC}$. It is very informative to compare how close to the true values each of these methods come when imputing the cellwise outliers. We report the performance of each imputed matrix $\mathbf{X}^{imp}$ as the root mean squared error of imputation (RMSEI) between the simulated uncontaminated matrix $\mathbf{X}$ and imputation of CRM and DDC:
\begin{equation}
\text{RMSEI}(\mathbf{X}^{imp}, \mathbf{X})=\sqrt{\frac{1}{np}\sum_{i=1}^{n}\sum_{j=1}^{p}(x_{ij}^{imp} - x_{ij})^2}
\end{equation}

Finally, it is an interesting question to investigate the quality of identification of cellwise outliers by CRM. The latter is reported as:
\begin{itemize}
    \item the {\em recall} (also called hit rate or true positive rate): of all cellwise outliers, how many have actually been detected as outliers (and therefore imputed) by CRM
    \item the precision: of all cells that were flagged as outliers by CRM, how many actually were cellwise outliers
\end{itemize}

\subsection{Results}

\begin{figure}[htbp]
\centering
\includegraphics[width=0.7\textwidth]{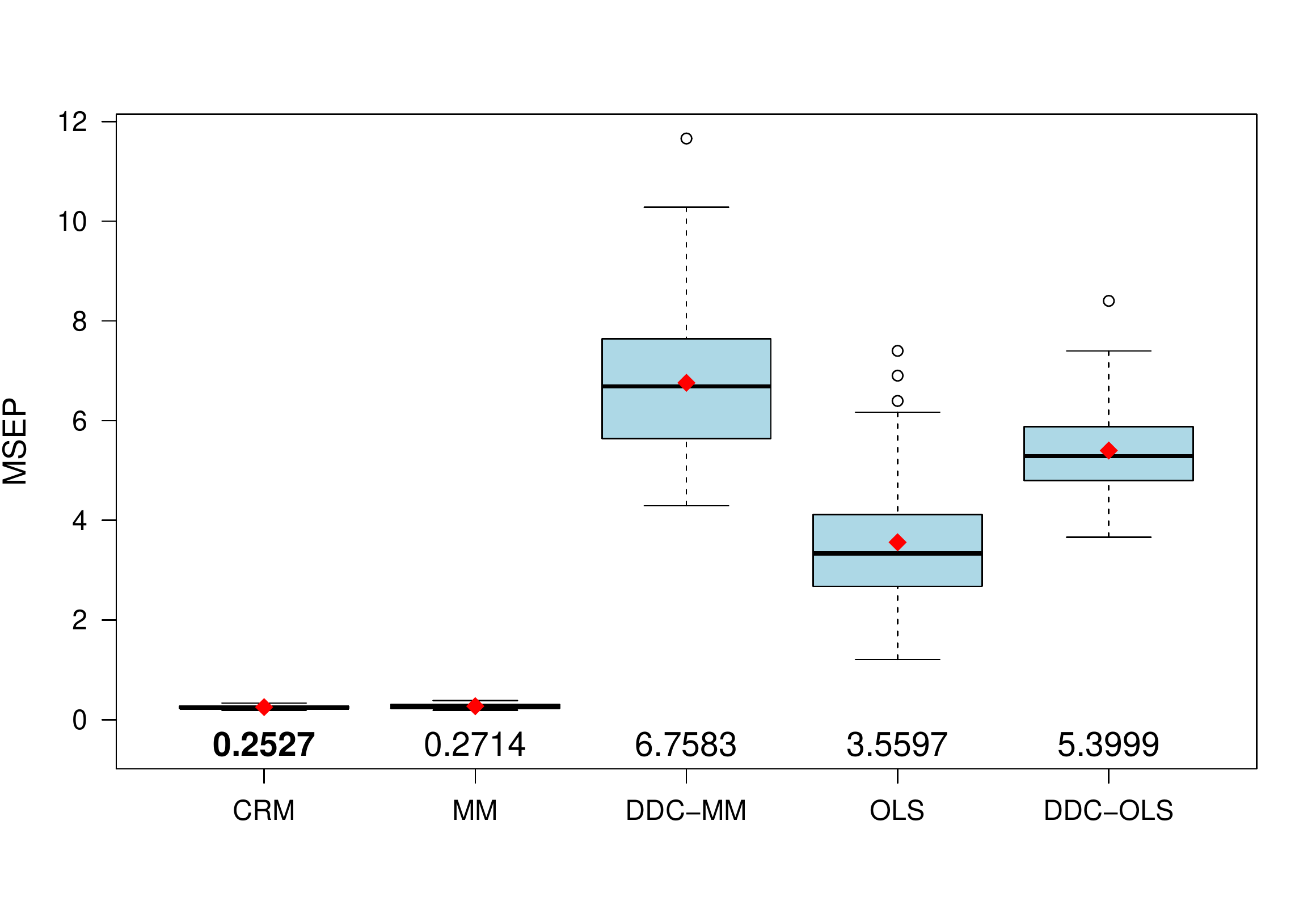}
~
\vspace*{-5mm}
\caption{\label{fig:boxplot_MSEP} Boxplot of MSEP for each of the regression methods. The average result for each method is printed at the bottom where the best result is shown in bold.}
\end{figure}

\begin{figure}[htbp]
\centering
\includegraphics[width=0.7\textwidth]{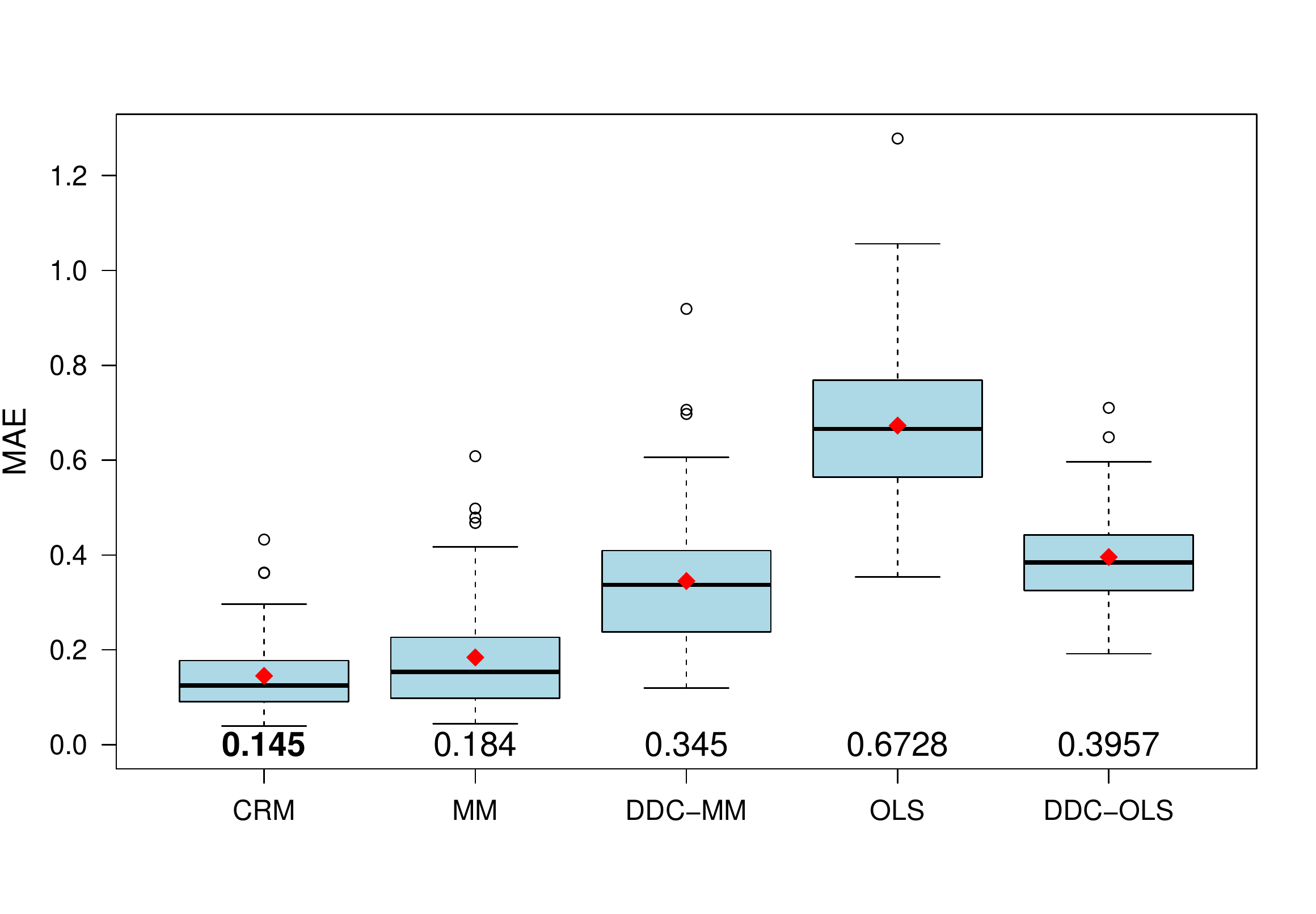}
~
\vspace*{-5mm}
\caption{\label{fig:boxplot_MAE} Boxplot of MAE for each of the regression methods.}
\end{figure}

\begin{figure}[htbp]
\centering
\includegraphics[width=0.7\textwidth]{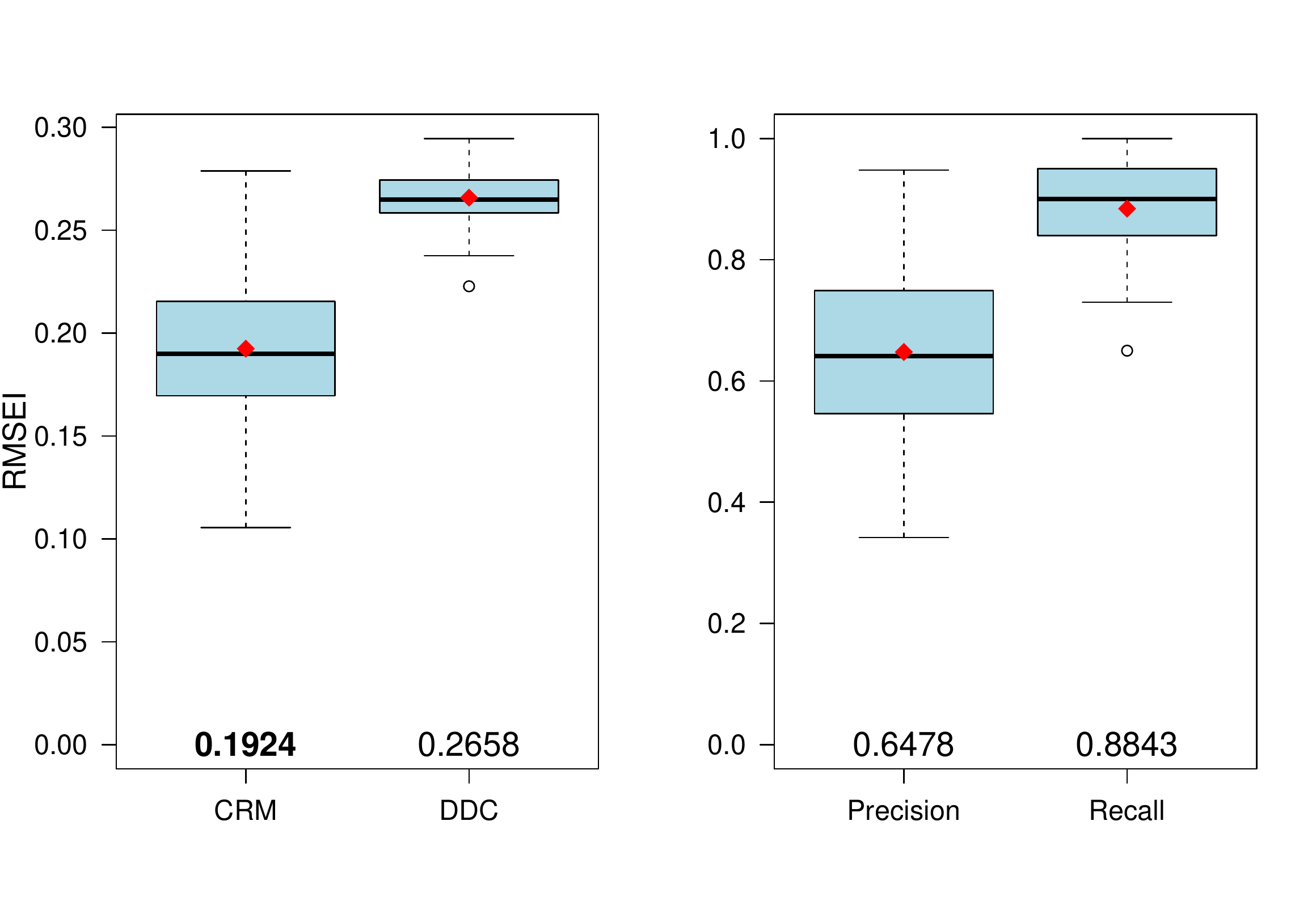}
~
\vspace*{-5mm}
\caption{\label{fig:boxplot_RMSEI_Precision_Recall}(Left) Boxplot of RMSEI for CRM and DDC. (Right) Precision and recall of detected cellwise outliers by CRM.}
\end{figure}

In what follows, the simulation results will be shown, illustrating the results according to the four evaluation criteria described in the previous section. 
Each result reported is the aggregate across hundred repeats. These aggregates are illustrated as boxplots  in Figures \ref{fig:boxplot_MSEP}, \ref{fig:boxplot_MAE}  and \ref{fig:boxplot_RMSEI_Precision_Recall}. The average result for each method is printed at the bottom and the best result is shown in bold.

From Figure \ref{fig:boxplot_MSEP}, one can derive that DDC based imputation as a preprocessing step to regression methods does not improve predictive performance. In fact, just applying least squares regression without DDC preprocessing predicts better. However, due to the presence of outliers, both robust regression methods, CRM and MM,
clearly outperform least squares. In terms of predictive power, 
Figure~\ref{fig:boxplot_MSEP} is reassuring in the sense that CRM performs equally well as the robust benchmark method, MM regression. 

Regarding bias in the regression coefficients, Figure \ref{fig:boxplot_MAE} shows that, not unexpectedly, the OLS regression coefficients are most biased in the presence of cellwise outliers. However, again applying DDC as a preprocessing step prior to either OLS or MM regression, performs much worse than casewise or cellwise robust M regression. 
The MAE for CRM regression is slightly smaller than that of plain MM regression,
and thus CRM has a slightly higher statistical efficiency than 
its casewise counterpart.

The subplots in Figure \ref{fig:boxplot_RMSEI_Precision_Recall} illustrate CRM's performance at imputing the true values for the cellwise outliers. In this respect, CRM beats DDC by a wide margin (left subplot). This is in line with CRM having a better predictive performance and a lower bias in the regression coefficients. Essentially, this tells us that much of CRM's superior performance shown in Figures \ref{fig:boxplot_MSEP} and \ref{fig:boxplot_MAE} can be attributed to it yielding a much more truthful intermediate imputation. Finally, the right panel in Figure \ref{fig:boxplot_RMSEI_Precision_Recall} shows that CRM imputation does well detecting and imputing cells that are cellwise outliers in reality, but it may sometimes impute a few too many.

On average it took 11.5 seconds to execute the CRM algorithm across the hundred repeats. The execution times were measured on an Intel core i5 with 2.7 GHz and 8 GB RAM.

\subsection{Simulation results for varying magnitude of contamination}
\label{sec:varyk}

\begin{figure}[htbp]
\centering
\includegraphics[width=0.49\textwidth]{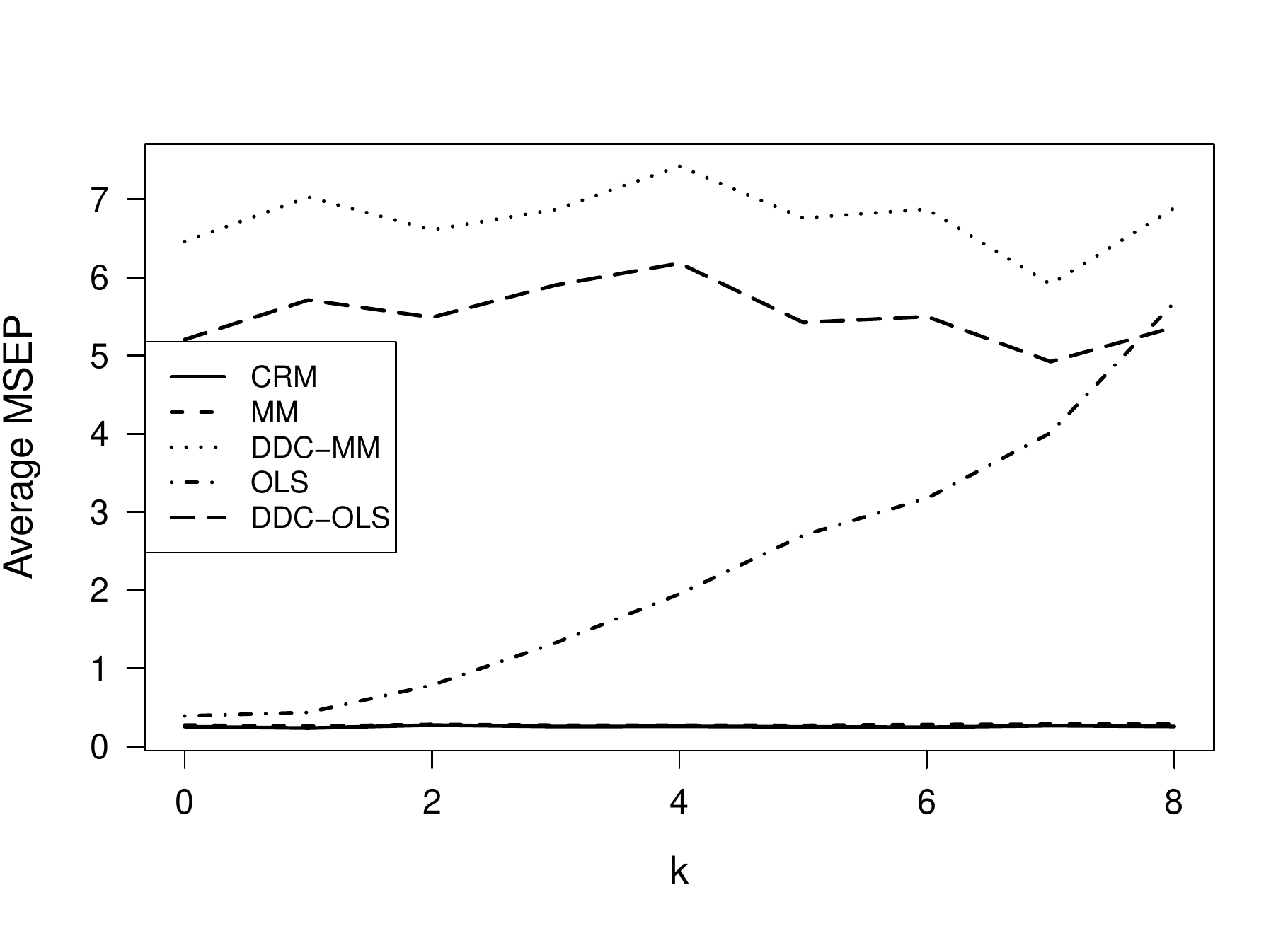}
\hfill
\includegraphics[width=0.49\textwidth]{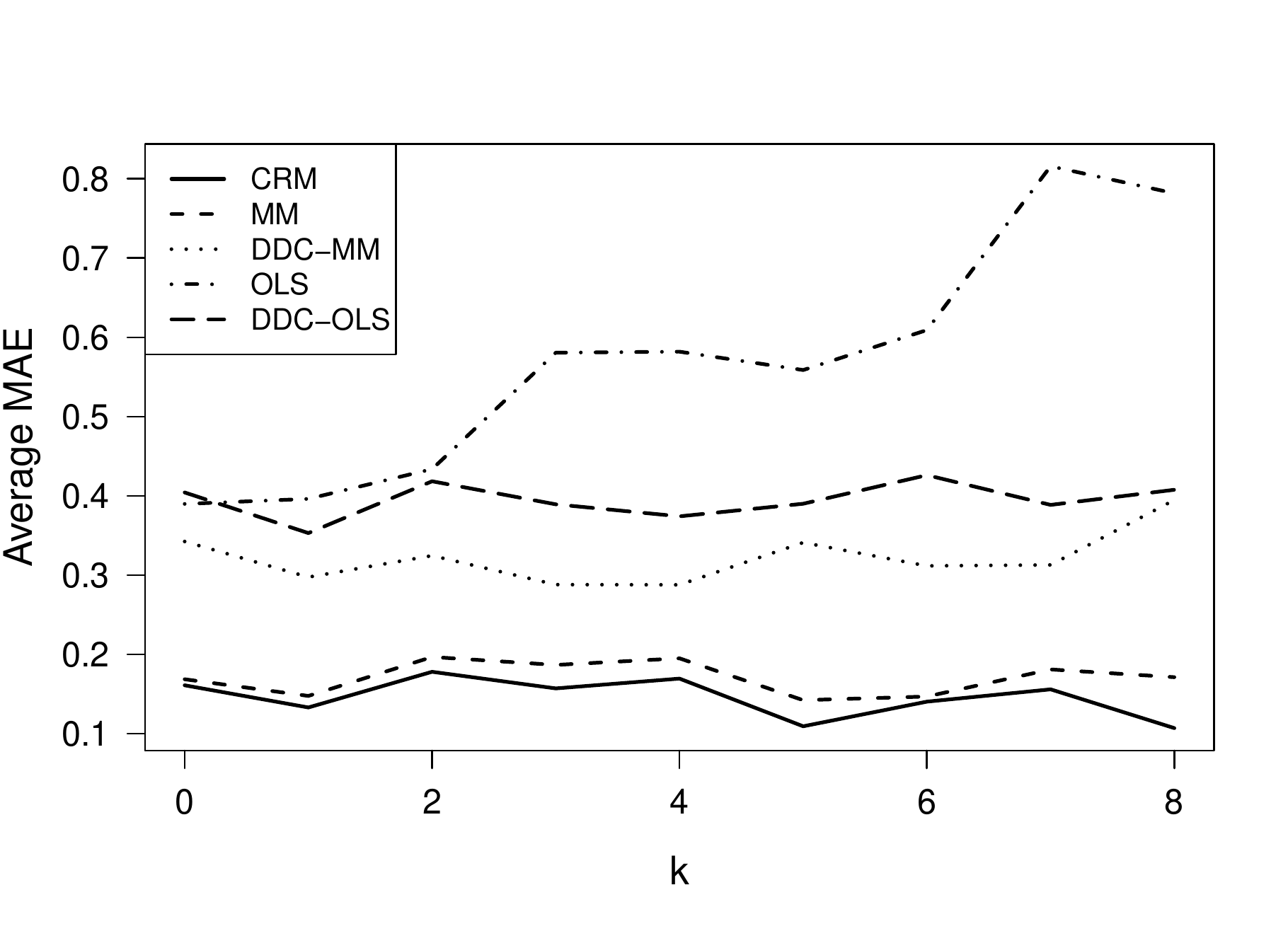}
~
\vspace*{-5mm}
\caption{\label{fig:MSEP_MAE} \textcolor{black}{Average MSEP (left) and 
MAE (right) for each of the regression methods for different values of $k$, controlling the magnitude of the outlyingness.}}
\end{figure}

\begin{figure}[htbp]
\centering
\includegraphics[width=0.5\textwidth]{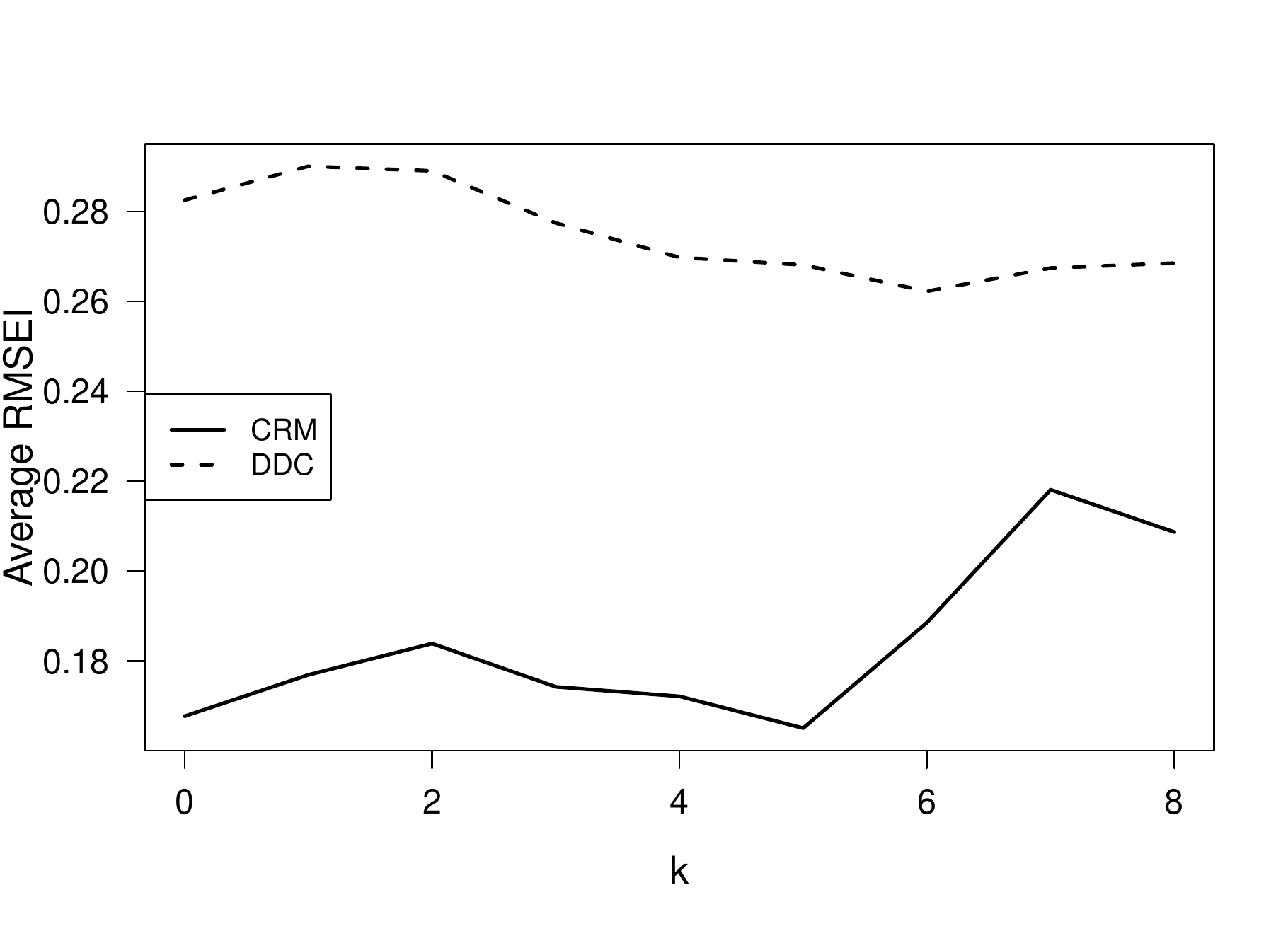}
~
\vspace*{-5mm}
\caption{\label{fig:RMSEI_curves} \textcolor{black}{Average RMSEI for CRM and DDC for different values of $k$, controlling the magnitude of the outlyingness.}}
\end{figure}

\begin{figure}[htbp]
\centering
\includegraphics[width=0.65\textwidth]{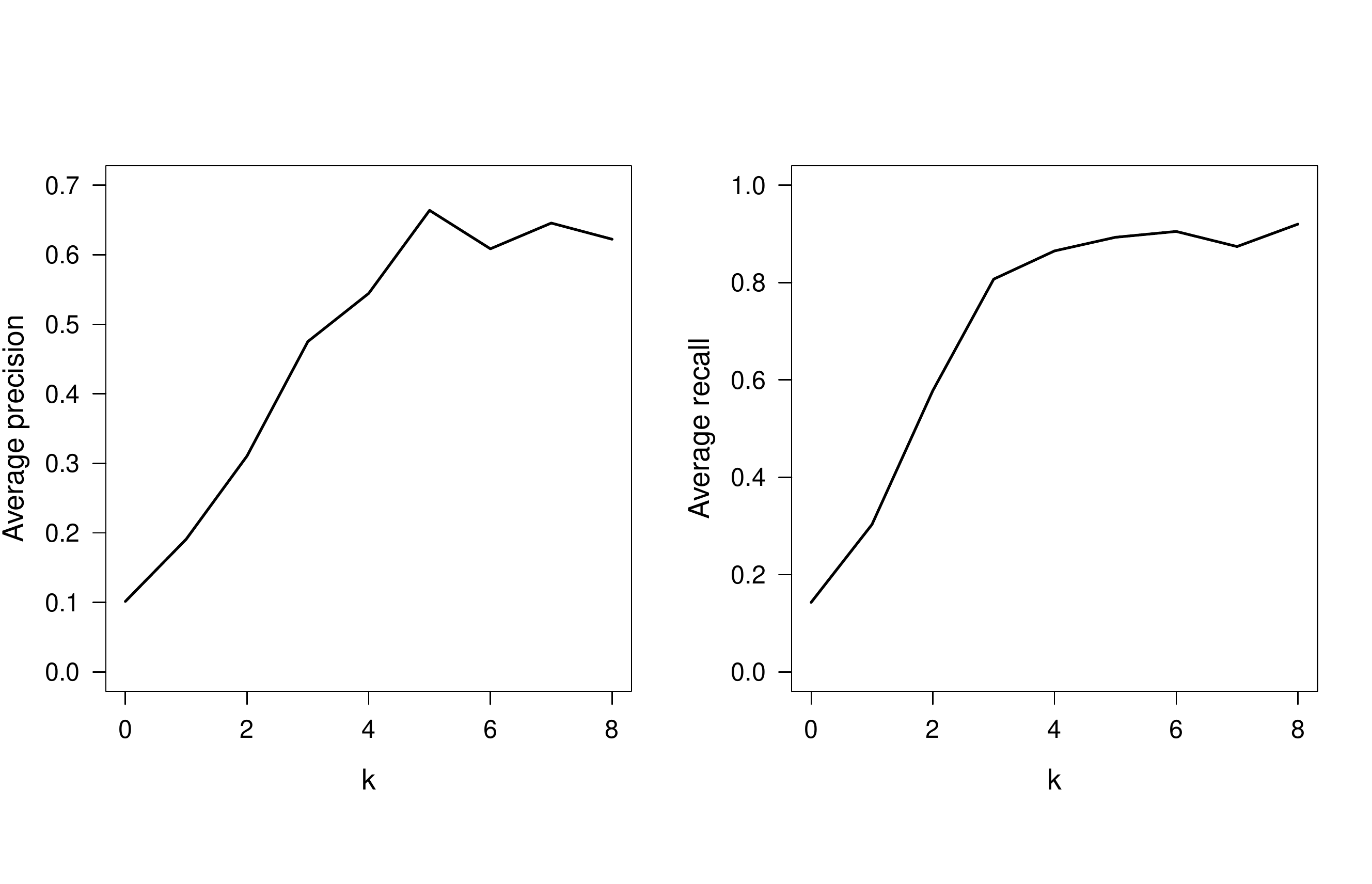}
~
\vspace*{-5mm}
\caption{\label{fig:precision_recall} \textcolor{black}{Average precision (left) and average recall (right) of detected cellwise outliers by CRM for different values of $k$, controlling the magnitude of the outlyingness.}}
\end{figure}

\textcolor{black}{
The way how the data cells were contaminated was described in Section~\ref{sec:contam},
where the magnitude of the contamination was controlled by the parameter $k$ as a 
multiple of the standard deviation of the respective variable. 
In this section we report simulation results when this parameter is varied
as $k\in\{0, 1, 2, \ldots, 8\}$, and thus we focus on the effect where the cells
are getting more and more extreme, starting from the variable average.
Figures~\ref{fig:MSEP_MAE}--\ref{fig:precision_recall} present the average
results across 10 simulation replications.
}

\textcolor{black}{
The results in Figure~\ref{fig:MSEP_MAE} indicate that already the situation
$k=0$, where the cells are replaced by the column means, creates outliers which 
are identified by CRM and MM regression, as well as by DDC. There is clearly a
strong effect on the prediction error (MSEP) and on the quality of the parameter
estimates (MAE) of the non-robust OLS estimator
if the contamination gets more extreme. All other estimators do not show a 
clear effect when $k$ is varied. However, Figures~\ref{fig:RMSEI_curves} and
\ref{fig:precision_recall} provide more insight: increasing $k$ leads to higher
precision and recall for CRM. In other words, CRM can only correctly identify the 
outlying cells (and not declare others as outliers)
if they are outlying clearly enough. With $k=0$, precision and 
recall are still very low. On the other hand, the resulting error from imputation
(RMSEI) for CRM is in the same range for $k$ varying from 0 to 6, and 
only for bigger $k$ it increases slightly, maybe due to an overfit effect,
but is still clearly smaller than that of DDC.
}

\subsection{Simulation results for breakdown}
\label{sec:breakdown}

\begin{figure}[htbp]
\centering
\includegraphics[width=0.6\textwidth]{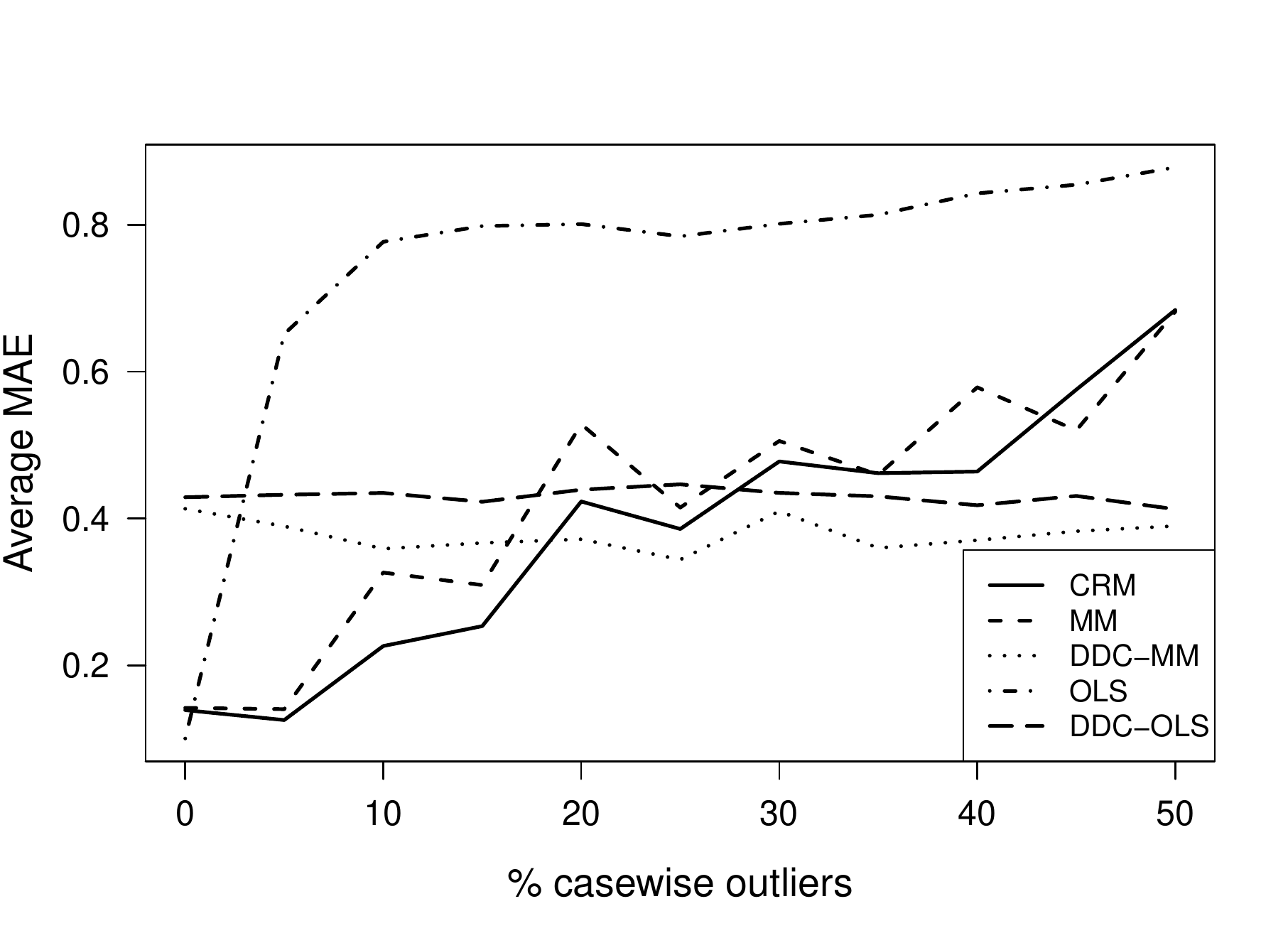}
~
\vspace*{-5mm}
\caption{\label{fig:breakdown} \textcolor{black}{Breakdown behavior of the different
regression methods: the fraction of contamination (10\% in each observation) is 
increased from 0 to 50\%.}}
\end{figure}

\textcolor{black}{
In the simulation settings considered so far, the fraction of contamination has
been fixed at 5\% (see Section~\ref{sec:contam}). It will now gradually be increased, while keeping all other parameters unchanged (here, $k=6$ is chosen again).
Thus, more and more out of the 400 observations are contaminated, where 
in each observation 10\% of the cells are randomly being picked for contamination.
Figure~\ref{fig:breakdown} shows the average of the results for MAE from all 
previously considered regression methods, starting from no contamination, up to 
50\% (in steps of 5\%). This reveals the breakdown behavior of the methods.
The OLS estimator is the most efficient when no contamination is present, but its
bias increases quickly when more and more contamination is added.
The DDC-based methods are almost not affected by the amount of contamination, because 
DDC is highly robust, but they lead to a comparably large bias for small percentages of contamination. MM-regression and CRM 
behave very similarly, but CRM seems to have a slightly smaller bias. Finally, note that the robustness 
behavior of both methods depends on tuning parameters; in case of CRM this depends
on the parameters of the Hampel function, which could be adjusted if necessary. 
}

\section{Real data example} \label{sec:Example}

%\subsection{Nutrients data}

The target of this analysis is to have a predictive model for cholesterol based on the nutrients contained in individual products. The data were taken from the Swiss nutrition data base 2015 \citep{SNT}. The original data set consists of nutrients on more than 40 components and 965 generic food products. We will focus on the first 193 products which do not contain any missing values and consider the variables in Table \ref{tab:nutrients_variables} where \verb|cholesterol| is the response variable. Since all of these 6 variables are skewed right, they were logarithmically transformed first.

\begin{table}[htbp]
\centering
\caption{Variables of the nutrients data.}
\label{tab:nutrients_variables}

\vspace*{2mm}
\begin{tabular}{l l}
\hline
 Variable & Description   \\ \hline
\verb|cholesterol| & cholesterol in milligram per 100g edible portion  \\
\verb|energy_kcal| & energy in kcal per 100g edible portion   \\
\verb|protein| & protein in gram per 100g edible portion  \\
\verb|water| & water in gram per 100g edible portion  \\
\verb|carbohydrates| & carbohydrates in gram per 100g edible portion  \\
\verb|sugars| & sugars in gram per 100g edible portion  \\
\hline
\end{tabular}
\end{table}

These data are a good example of data where one would expect the cellwise robust estimation technique to outperform the casewise one. While it is variable to assume a multivariate interplay between these variables in real life systems, these variables are measured independently. Moreover, biological effects can generate deviating behavior independently. It is therefore plausible to assume the necessity for multiple regression and corresponding multivariate effects taking place between the inputs, yet the mechanisms that generate outliers can be assumed to be largely independent. 

\begin{figure}[htbp]
\centering
\includegraphics[width=1\textwidth]{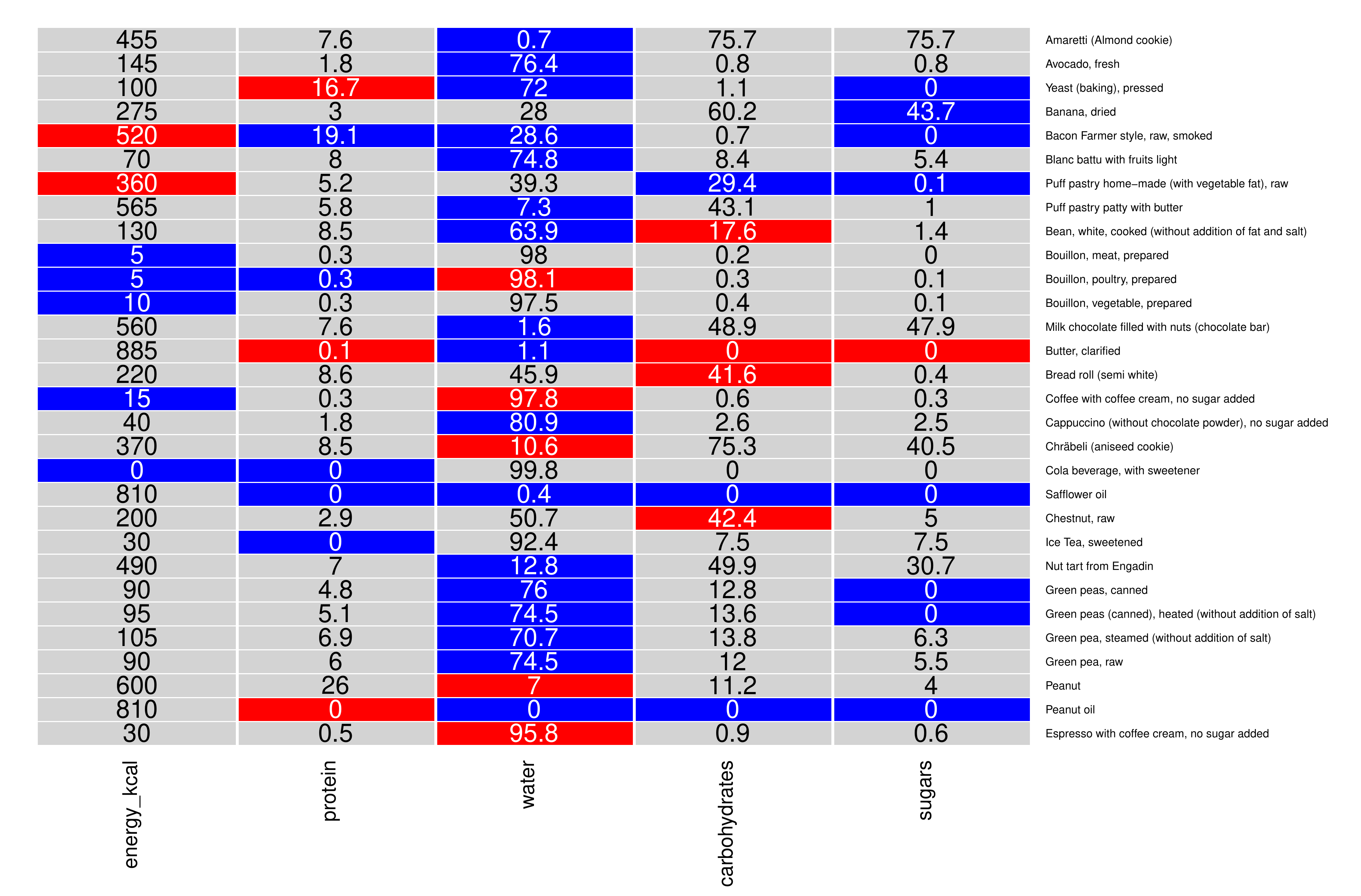}
~
\vspace*{-5mm}
\caption{Heatmap of outliers detected by CRM in the nutrients data. The red and blue boxes indicate the contaminated cells.}
\label{fig:nutrients_heatmap}
\end{figure}

\begin{figure}[htbp]
\centering
\includegraphics[width=1\textwidth]{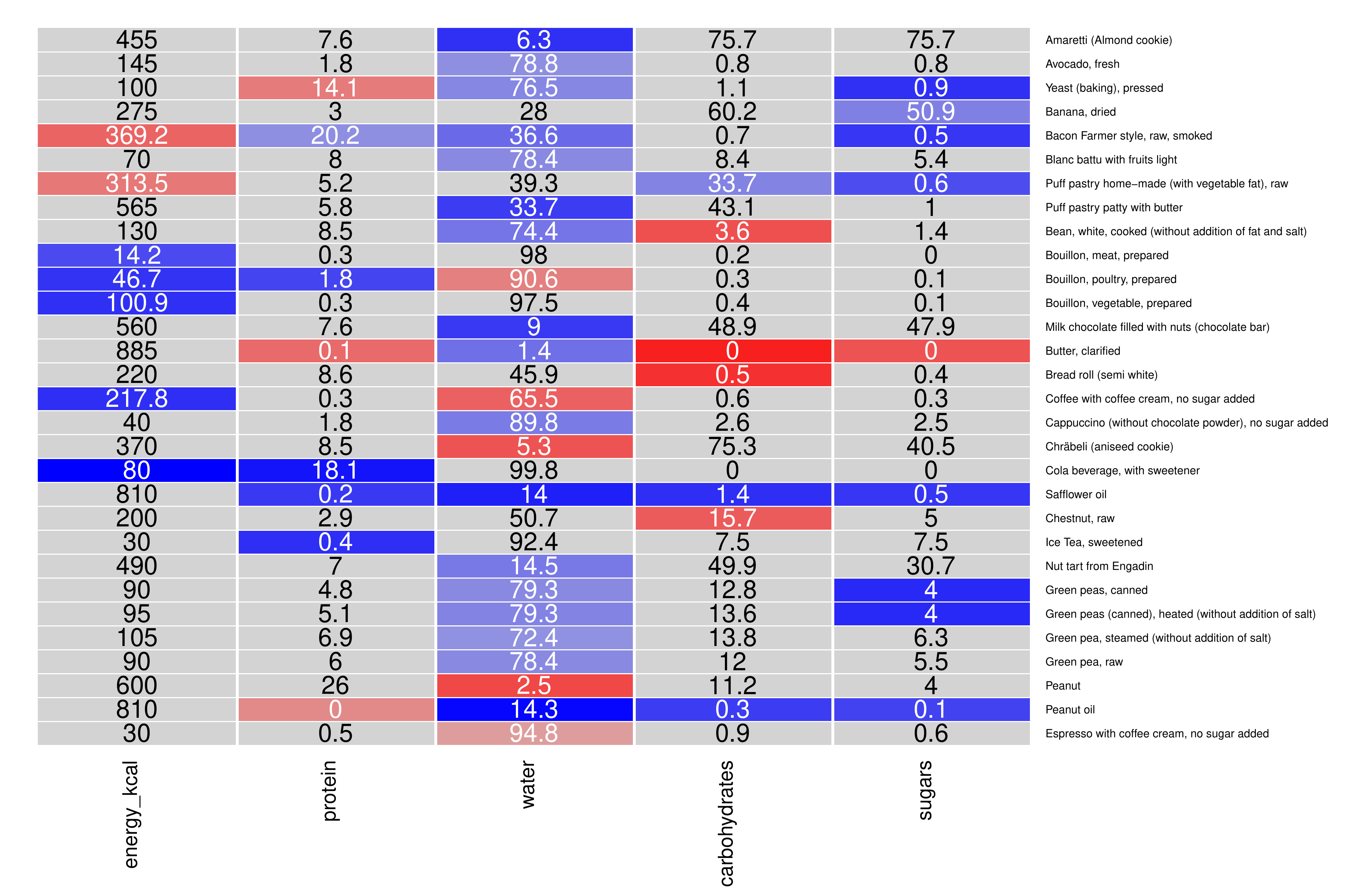}
~
\vspace*{-5mm}
\caption{Heatmap of outliers detected by CRM in the nutrients data. The red and blue boxes indicate the cells imputed by CRM, \textcolor{black}{and higher
saturation of the color refers to bigger differences to the original data}.}
\label{fig:nutrients_heatmap_imputed}
\end{figure}

The estimates of the regression coefficients by CRM are given in Table \ref{tab:nutrients_coefficients}. It took around 1 second to apply the CRM algorithm on this dataset. CRM indicates 30 out of 193 food products as casewise outliers having at least one contaminated outlying cell. The results are plotted as a heatmap in Figure \ref{fig:nutrients_heatmap}, where the 30 outliers are represented as rows and each contaminated/outlying cell is shown as a colored box. The anomalous cells, whose values are deviating either upwards or downwards, are colored red or blue, respectively. 
It can be seen that some food products have a lot of anomalous nutrient values, whereas others only have an atypical value for a few cells. Figure \ref{fig:nutrients_heatmap_imputed} contains the nutritional data of the 30 anomalous food products where the deviating cells have been imputed by CRM. The blue cells are replaced with larger values while the red cells are imputed with smaller values (according to CRM). \textcolor{black}{A more saturated color refers to bigger differences
between the imputed and the original data values.}

\begin{table}[htbp]
\centering
\caption{Estimated regression coefficients of the nutrients data by CRM.}
\label{tab:nutrients_coefficients}

\vspace*{2mm}
\begin{tabular}{l r}
\hline
Variable & Estimated coefficient  \\ \hline
\verb|(Intercept)| & -33.73173  \\
\verb|log.energy_kcal| &  3.62970  \\
\verb|log.protein| &  0.98341  \\
\verb|log.water| & 3.78561  \\
\verb|log.carbohydrates| & 0.05336 \\
\verb|log.sugars| & -0.10999 \\
\hline
\end{tabular}
\end{table}

\begin{figure}[htbp]
\centering
\includegraphics[width=0.7\textwidth]{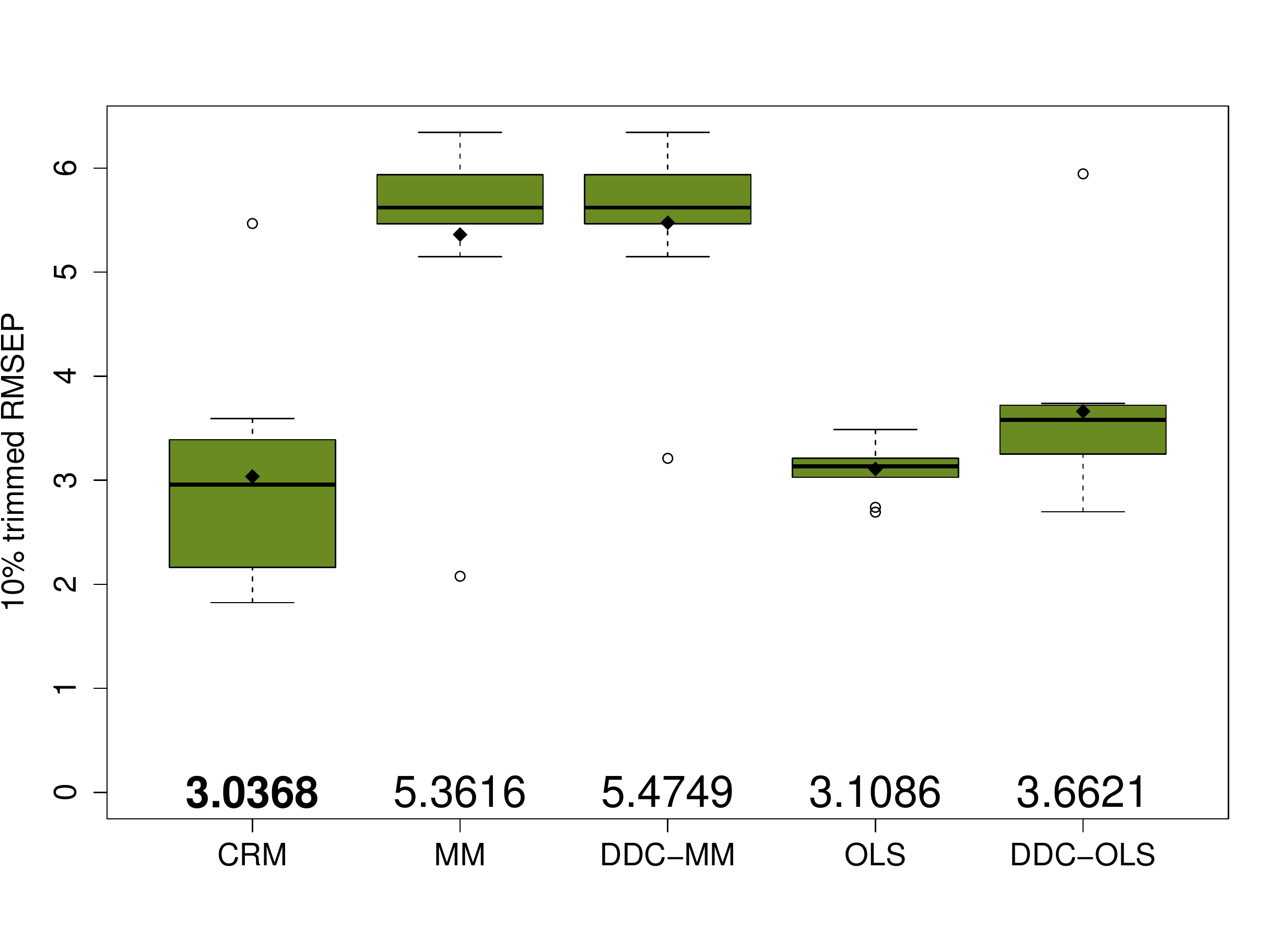}
~
\vspace*{-5mm}
\caption{Boxplot of $10\%$ trimmed RMSEP values from 10-fold cross validation for each of the regression methods.}
\label{fig:nutrients_crossvalidation2}
\end{figure}

A 10-fold cross validation is conducted on the nutrients data set, using 
\textcolor{black}{CRM as well as OLS and MM regression, also in the version where
outliers are first replaced with the DDC method (DDC-OLS and DDC-MM).}
Figure \ref{fig:nutrients_crossvalidation2} shows that CRM clearly outperforms MM regression 
\textcolor{black}{as well as the DDC-MM method}
in terms of the $10\%$ trimmed root mean squared error of prediction (RMSEP).
\textcolor{black}{
Note that trimming makes the evaluation measure robust against the 
outliers~\citep{maronna2019robust}.
As Figures \ref{fig:nutrients_heatmap} and \ref{fig:nutrients_heatmap_imputed} show,
CRM only imputes the cellwise outliers, which often just amount to a single cell per case. This implies that CRM can process up to 80\% more relevant information for the outlying cases when compared to MM regression, that downweights these entire cases, some of them even to zero. Because CRM retains more uncontaminated information, it allows to make more accurate predictions.}
\textcolor{black}{There is, however, only a smaller difference between CRM 
and OLS or DDC-OLS. It seems 
that the outliers in the explanatory variables are not acting as bad leverage points,
and correcting them by DDC even leads to a slightly poorer prediction performance 
for OLS regression. The imputation by CRM on the other hand seems much more 
appropriate. Even though the predictive performance is comparable to OLS, 
CRM has the advantage that it produces heatmaps of the outlying cells and the 
values that they have been imputed with, see Figure~\ref{fig:nutrients_heatmap_imputed}}.
These can be of great value to the practitioner, since they allow to analyze which cells deviate per case and understand the outlier generating mechanism(s).

\section{Conclusions and outlook} \label{sec:Outro}

Cellwise robust M regression has been introduced as a regression method that is robust to vertical outliers and both cellwise and casewise and leverage points. Intrinsically, the method detects cells that are deviating {\em with respect to the linear model} and imputes them with more model consistent values. While CRM may not be the first method to detect deviating cells, it is the first to do so in a model consistent way for a linear model. This offers the practitioner a combined advantage of having a robust fit that reliably fits the majority of the data and producing a heatmap of suspected deviating cells, as well as model based imputations. Compared to casewise robust estimators, CRM will retain a larger fraction of uncontaminated data cells. Depending on the data, this fraction can be substantially larger. Therefore, the resulting fit is closer to the underlying model that generates the data, and the procedure is more efficient. 

A simulation study has shown that CRM can generally be assumed to perform on par with a casewise robust estimator in terms of predictive power; an example prediction cholesterol from other nutrients has shown, though, that real life cases {\em can} be found for which CRM does outperform the casewise robust estimator. The simulation study has also shown that when a linear model can be assumed to have generated the data, detecting deviating cells and imputing them by CRM is a much better idea than using technique agnostic of the linear model, such as DDC, prior to regression. 
\textcolor{black}{Also alternatives to DDC, for example kNN-based imputation methods,
would suffer from the same problem that they
ignore the underlying model.}
Finally, the simulation study has highlighted as well that CRM is slightly more efficient than MM at estimating individual regression coefficients. 

 On a casewise basis, there is widespread consensus that outliers are only outlying with respect to a model, and therefore need to be detected by robust estimators for the corresponding model. Few experts will assume that there exists a generic {\em data cleaning} estimator that can generically detect all outliers, regardless of which model is being estimated. This has led to over forty years of research on robust statistics, that has produced robust counterparts for virtually the entire arsenal of classical statistical estimators. However, on a cellwise basis, the few approaches in the literature, such as DDC, seem to be going for generically detecting cellwise outliers. DDC has practical merit: it can be applied to data sets that contain over 50\% of casewise outliers, for which CRM would break down. That said, the majority of data sets containing cellwise outliers do not contain them in a majority of cases. In the latter situation, we argue that it is better to construct a model consistent cellwise robust estimator. The simulation study and example have corroborated this claim.   
 
The introduction of CRM is a first step that opens the door to an entire class of model consistent cellwise robust estimators. Just like for casewise robust statistics, we would like to open up the field and see development starting on other cellwise robust statistics, such as cellwise robust principal component analysis, cellwise robust partial least squares, or cellwise robust canonical correlation analysis, just to name a few. 
 
Another topic for further research would be inference. It has been shown that Wald type inference can be applied to MM-estimators \citep{Koller}; these results have been incorporated into the \texttt{R} package \texttt{robustbase}. It will be a promising topic of research to extend these results to the cellwise setting. Note that while analytical results on inference for the CRM parameters are still to be generated, the practitioner can always resort to the robust bootstrap, which can be computed fast \citep{FRB}.  

\section*{Acknowledgments}

\textcolor{black}{The authors are grateful to two anonymous reviewer. Their suggestions
and remarks led to a substantially improved manuscript.}

%\section*{References}
 
\bibliographystyle{plainnat}
%\bibliography{ref}

\end{document}